\documentstyle[emulateapj,flushrt,tighten]{article} 
\input{psfig.sty}
\def\kms{\,{\rm km\,s^{-1}}}

\def\msun{\,{\rm M_\odot}}

\def\etal{{et al.\ }}

\newcommand\beq{\begin{equation}}
\newcommand\eeq{\end{equation}}
\newcommand{\ba}{\begin{eqnarray}}
\newcommand{\ea}{\end{eqnarray}}
\def\spose#1{\hbox to 0pt{#1\hss}}
\def\lta{\mathrel{\spose{\lower 3pt\hbox{$\mathchar"218$}}
      \raise 2.0pt\hbox{$\mathchar"13C$}}}
\def\gta{\mathrel{\spose{\lower 3pt\hbox{$\mathchar"218$}}
      \raise 2.0pt\hbox{$\mathchar"13E$}}}

\lefthead{Volonteri, Madau, \& Haardt}
\righthead{Formation of Galaxy Stellar Cores}
\makeatletter

\newenvironment{figurehere}
  {\def\@captype{figure}}
  {}
\makeatother

\begin{document}
\submitted{}

\title{The formation of galaxy stellar cores by the hierarchical merging of 
supermassive black holes} 

\author{Marta Volonteri\altaffilmark{1}, Piero Madau\altaffilmark{1}, \& 
Francesco Haardt\altaffilmark{2}}

\altaffiltext{1}{Department of Astronomy \& Astrophysics, University of California, 
Santa Cruz, CA 95064.}
\altaffiltext{2}{Dipartimento di Scienze, Universit\`a dell'Insubria/Sede di Como, 
Italy.} 

\slugcomment{ApJ, in press}

\begin{abstract}
We investigate a hierarchical structure formation scenario in which galaxy 
stellar cores are created from the 
binding energy liberated by shrinking supermassive black hole (SMBH) binaries.
The binary orbital decay heats the surrounding stars, eroding a preexisting
stellar cusp $\propto r^{-2}$. We follow the merger history of dark matter 
halos and associated SMBHs via cosmological Monte Carlo realizations of the 
merger hierarchy from early times to the present in a $\Lambda$CDM cosmology.
Massive black holes get incorporated through a series of mergers 
into larger and larger halos, sink to the center owing to dynamical friction, 
accrete a fraction of the gas in the merger remnant to become supermassive,
and form a binary system. Stellar dynamical processes drive the binary to 
harden and eventually coalesce. A simple scheme is applied in which the 
loss cone is constantly refilled and a constant density core forms 
due to the ejection of stellar mass. We find that a model in which the 
effect of the hierarchy
of SMBH interactions is cumulative and cores are preserved during galaxy
mergers produces at the present epoch a correlation between the 
`mass deficit'
(the mass needed to bring a flat inner density profile to a $r^{-2}$ cusp)
and the mass of the nuclear SMBH, with a normalization and slope 
comparable to the observed 
relation. Models in which the mass displaced by the SMBH binary
is replenished after every major galaxy merger appear instead to 
underestimate the mass 
deficit observed in `core' galaxies. 
\end{abstract}

\section{Introduction}

The strong link observed between the masses of supermassive black holes 
(SMBHs) residing at the center of most nearby galaxies and the 
gravitational potential wells that host them (Ferrarese \& Merritt 2000; 
Gebhardt \etal 2000; Ferrarese 2002) suggests a fundamental 
mechanism for assembling black holes (BHs) and forming spheroids in 
galaxy halos. In popular cold dark matter (CDM) `bottom-up' cosmogonies, 
the host galaxies of SMBHs experience multiple mergers during 
their lifetime (Kauffmann \& Haehnelt 2000; Menou, Haiman, \& Narayanan 
2001),
with those between comparable-mass systems (`major mergers') expected to 
result in the formation of elliptical galaxies
(Barnes 1988). If the hierarchical build-up of SMBHs traces far 
up in the dark halo merger tree, binary BH systems may 
be widespread (Volonteri, Haardt, \& Madau 2003, hereafter Paper I). 
As galaxies hosting nuclear BHs merge, the holes will sink to the center of the
merger remnant owing to dynamical friction from field particles, and form 
a bound pair. 

If was first proposed by Ebisuzaki, Makino, \& Okumura (1991) that 
the heating 
of the surrounding stars by a decaying SMBH pair would create a low-density 
core out of a preexisting cuspy stellar profile. In a stellar background a 
`hard' binary shrinks by capturing the stars that pass 
close to the holes and ejecting them at much higher velocities, a super-elastic
scattering process that depletes the 
nuclear region. Rapid coalescence eventually ensues due to the
emission of gravitational radiation. 
Observationally, there is clear evidence in early-type galaxies for a 
systematically different 
distribution of surface brightness profiles, with faint ellipticals showing 
steep power-law profiles (cusps), while bright ellipticals have much 
shallower stellar cores (Lauer \etal 1995; Gebhardt \etal 1996; Faber 
\etal 1997; Ravindranath \etal 2001). `Core' galaxies have high 
velocity dispersions and exhibit a definite break in the brightness profile
at some radius that scales with luminosity as $\propto L_V^{1.15}$ 
(Faber \etal 1997). Detailed N-body simulations have confirmed the 
cusp-disruption effect of a hardening BH binary (Makino \& Ebisuzaki 1996;
Quinlan \& Hernquist 1997; Milosavljevic \& Merritt 2001), but have shed 
little light on why bright ellipticals have lower central concentrations 
than do faint ellipticals. 

The role of binaries in shaping the central structure of galaxies can be
best
understood within the framework of a detailed model for the hierarchical 
assembly of SMBHs over 
cosmic history (Paper I; Haehnelt \& Kauffmann 2002), particularly if the damage 
done to a stellar cusps by decaying BH pairs is cumulative and nuclear cores are 
preserved during galaxy mergers. In this paper we study the 
effects of hierarchical mergers of halo$+$SMBH systems on the inner density 
profiles of galaxies using the machinery for following the growth and dynamics
of SMBHs developed 
in Paper I. We show that stellar cusps can be efficiently destroyed over 
cosmic time by decaying SMBH binaries if stellar dynamical processes are able
to shrink the binary down to a separation $\lta 10\%$ of the separation
at which the binary becomes `hard'. More massive halos have more massive
nuclear BHs and experience more merging events than less massive galaxies:
hence they suffer more from the eroding action of binary SMBHs and have 
larger cores.  

\section{Assembly and growth of SMBH\lowercase{s}}

We briefly summarize here the main features of our scenario for the
hierarchical growth of SMBHs in a $\Lambda$CDM cosmology (see Paper I for a 
thorough discussion). The merger history of 220 parent halos with present-day 
masses in the range $10^{11}<M_0<10^{15}\,\msun$ is tracked backwards with a 
Monte Carlo algorithm based on the extended Press-Schechter formalism. 
Compared to Paper I, we use an improved version where the most massive halos 
are broken up into as many as 280,000 progenitors by $z=20$. We adopt a 
two-component model for galaxy halos. The dark matter is distributed 
according to a NFW profile (Navarro, Frenk, \& White 1997),
\begin{equation}
\rho_{\rm DM}(r)={M\over 4 \pi r (r+r_{\rm vir}/c)^2 f(c)},
\label{eqn:nfw1}
\end{equation}
where $r_{\rm vir}$ is the virial radius, $c$ is the halo concentration 
parameter and 
\begin{equation}
f(c)=\ln(1+c)-{c\over 1+c}.
\end{equation}
Following Bullock \etal (2001), the mean
concentration is assumed to scale with halo mass $M$ and redshift of 
collapse $z$ as 
\beq 
\bar c={9\over 1+z}\left({M\over 8\times 10^{12} 
h\msun}\right)^{-0.14}.
\eeq
Individual halos have a log-normal distribution with dispersion about the mean 
$\Delta (\log~c)=0.18$. During the merger of two halo$+$BH systems of 
comparable masses, dynamical friction against the dark matter background  
drags in the satellite hole towards the center of the newly merged system,
leading to the formation of a bound BH binary in the violently relaxed stellar
core. The dynamical friction timescale depends on the orbital parameters of 
the infalling satellite, which we take from van den Bosch \etal (1999).  
At late epochs, most of the BH pairs have unequal masses, with 
mass ratios ranging between 10\% and 20\% (Paper I).

The subsequent evolution of the binary is determined by the initial 
central stellar distribution. We model this as a singular isothermal 
sphere (SIS) with one-dimensional velocity dispersion $\sigma_*$ and 
density 
\beq 
\rho_*(r)={\sigma_*^2\over 2\pi G r^2}.
\eeq
The stellar velocity dispersion is
related to the halo circular velocity $V_c$ at the virial radius following
Ferrarese (2002), 
\beq
\log V_c=(0.88\pm 0.17)\log \sigma_*+(0.47\pm 0.35).
\eeq
We truncate the stellar SIS at $0.16\,r_{\rm vir}$ in order for the total 
stellar mass fraction to equal the universal baryon fraction, 
$\Omega_b/\Omega_M$.  In our model pregalactic `seed' holes form with 
intermediate masses ($m_\bullet=150\,\msun$) in (mini)halos 
collapsing at $z=20$ from rare 3.5-$\sigma$ peaks of the primordial density 
field (Paper I; Madau \& Rees 2001). The assumed `bias' assures that 
almost all halos above $10^{11}\,\msun$ actually host a BH at all epochs. 
We found little
change in the $z<5$ results in a test model case with $m_\bullet=1000\,\msun$. 
In each major merger the more massive hole accretes at the 
Eddington rate a gas mass that scales with the fifth power of the circular 
velocity of the host halo,
\begin{equation}
\Delta m_{\rm acc}=3.6\times 10^6\,\msun~{\cal K}\,V_{c,150}^{5.2},
\label{macc_eq}
\end{equation}
where $V_{c,150}$ is the circular velocity of the merged system in units of
150 $\kms$.\footnote{In eq. (\ref{macc_eq}) a typographical error that 
appeared in the coefficient of eq. (13) in Paper I has been corrected.}   
The normalization is fixed a posteriori in order
to reproduce the observed local $m_{\rm BH}-\sigma_*$ relation (Ferrarese
2002).
{\it The present-day mass density of nuclear SMBHs accumulates mainly 
via gas accretion, with BH-BH mergers playing only a secondary role}. This 
model was shown in Paper I to reproduce remarkably well the observed 
luminosity 
function of optically-selected quasars in the redshift range $1<z<5$.

If the merging timescales of SMBH binaries are
longer than the characteristic timescale between major galaxy mergers,
then interactions between the binary and new infalling holes will be likely.
In Paper I we discussed a scenario where binaries decay  
efficiently both as a result of mass ejection from a cuspy stellar density 
profiles and, at very high redshifts, due to triple BH interactions. 
In the next section we expand upon Paper I and describe a scheme for 
generating low-density cores out of a preexisting stellar cusps from the 
binding energy liberated by shrinking SMBH binaries.

\section{Dynamical evolution of SMBH binaries}

Consider a binary with BH masses $m_1\ge m_2$ and
semimajor axis $a(t)$ in an isotropic background of stars of mass $m_*\ll 
m_2$ and density $\rho_*(r)$. 
The binary will initially shrink by dynamical friction from 
distant stars acting on each BH individually. But as the binary separation 
decays, the
effectiveness of dynamical friction slowly declines because distant encounters 
perturb only the binary center of mass but not its semimajor axis. The BH pair 
then hardens via three-body interactions, i.e., by capturing and ejecting at 
much higher velocities the stars passing by within a distance $\sim a$ 
(`gravitational slingshot'). The 
system becomes hard when the value of $a$ falls below 
\beq
a_h={Gm_2\over 4\sigma_*^2}
\eeq
(Quinlan 1996).
We assume that the `bottleneck' stages of the (bound) binary shrinking 
occur for 
separations $a<a_h$; during a galactic merger, after a dynamical friction 
timescale, we place the BH pair at $a_h$ and let it evolve.   

The hardening of the binary modifies the stellar density profile,
removing mass interior to the binary orbit, 
depleting the galaxy core of stars, and slowing down further hardening. 
If ${\cal M}_{\rm ej}$ is the stellar mass ejected by the BH pair, 
the binary evolution and its effect on the galaxy core are determined by 
two dimensionless quantities: the hardening rate
\beq
H={\sigma_*\over G\rho_*}{d\over dt}{1\over a},
\label{eqH}
\eeq
and the mass ejection rate
\beq
J={1\over (m_1+m_2)}\,{d{\cal M}_{\rm ej}\over d\ln(1/a)}.
\label{eqJ}
\eeq
A third dimensionless quantity, the eccentricity growth rate, was shown 
by Quinlan (1996) to be unimportant and will be neglected here. The quantities
$H$ and $J$ can be found from scattering experiments that treat the 
star-binary encounters one at a time. Following Quinlan (1996), we take 
$H=15$ independent of $a$ (for $a<a_h$) and of the binary
mass ratio. The quantity $J$ is instead a function of binary 
separation and has some dependence on $m_1/m_2$. We inter/extra-polate 
Quinlan's numerical results to obtain $J(a)$ for different mass ratios (cf.
Paper I). The merger timescale is computed adopting a simple 
semi-analytical scheme that qualitatively reproduces the evolution observed
in N-body simulations (Merritt 2000; Paper I). We assume that 
the stellar mass removal creates a core of radius $r_c$ and constant 
density $\rho_c\equiv \rho_*(r_c)$, so that the total mass ejected as
the binary shrinks from $a_h$ to $a$ can be written as
\begin{equation}
{\cal M}_{\rm ej}= {2\sigma_*2\over G}(r_c-r_i)+M_i-M_c
= {4\over 3}{\sigma_*^2 (r_c-r_i)\over G}, 
\label{rcore}
\end{equation}
where 
$M_c=4\pi\rho_cr_c^3/3$, $M_i=4\pi\rho_ir_i^3/3$, 
$r_i=r_c(t=0)$ is the radius of the (preexisting) core when 
the hardening phase starts at $t=0$, and $\rho_i\equiv \rho_*(r_i)$. 
The core radius then grows as 
\begin{equation}
r_c(t)=r_i+{3\over4 \sigma_*^2}G(m_1+m_2)\int_{a(t)}^{a_h}{\frac{J(a)}{a}\,da}.\label{rc}
\end{equation}
The binary separation quickly falls below $r_c$ and subsequent evolution 
is slowed down due to the declining stellar density, with a hardening time,
\beq
t_h=|a/\dot a|={2\pi r_c(t)^2\over H\sigma_*a},
\eeq
that becomes increasingly long as the binary shrinks. The mass ejected
increases approximately logarithmically with time, and the binary `heats' 
background stars at radii $r_c\gg a$. If the hardening continues sufficiently far,
gravitational radiation losses finally take over, and the two BHs coalesce 
in less than a Hubble time.\footnote{As in Paper I, the final mass $m_{\rm BH}$ of the SMBH 
formed after coalescence assumes the entropy-area relation for BHs (maximally 
efficient radiative merging): $m_{\rm BH}^2=m_1^2+m_2^2$.}
In the scheme for core creation investigated here, at all redshift $z<5$
the binary hardening timescale is always less than the time it takes 
the `satellite' halo$+$BH system to sink to the center of the more massive 
progenitor due to dynamical friction against the dark matter background.

The above relations assume that the stellar velocity dispersion remains 
constant during the hardening of the binary. This is a reasonable 
assumption as the two-body stellar relaxation timescale
\beq
t_r=0.34\frac{\sigma^3_*}{G^2 m_* \rho_c \ln\Lambda}\,=
7\,{\rm Gyr}\left(\frac{r_c}{\rm pc}\right)^2\left(\frac{10}{\ln \Lambda}
\right)\sigma_{150},
\label{trelax}
\eeq
is typically much longer than the hardening time. 
Here $\Lambda$ is of order the total number of stars, $m_*$ is the 
typical stellar mass, and $\sigma_{150}$ is the stellar velocity dispersion 
in units of $150\,\kms$. 
More importantly, we have neglected the depopulation of the loss cone, since 
it is the total stellar 
density that is allowed to decrease following equation (\ref{rc}), not the 
density of low angular momentum stars. The effect of 
loss-cone depletion (the depletion of low-angular momentum stars that get close enough 
to extract energy from a hard binary) is one of the major uncertainties in computing 
the merger timescale, and makes it difficult to construct viable merger scenarios for BH binaries.
The wandering of the binary center of mass from the galaxy center induced by 
continuous 
interactions with background stars (Quinlan \& Hernquist 1997), the large supply of low-angular 
momentum stars in significantly flattened or triaxial galaxies (Yu 2002), the presence of a
third BH (Valtonen \etal 1994; Blaes, Lee, \& Socrates 2002; Paper I), 
the loss 
of orbital angular momentum to a gaseous disk (Gould \& Rix 2000; Armitage \& Natarajan 
2002), and the randomization of stellar orbits due to the infall of small satellites
(Zhao, Haehnelt, \& Rees 2001) will all mitigate to some extent the problems associated with 
loss-cone depletion (which may ultimately cause the binary to stall) and 
help the binary merge.

\section{Results}

The ability of SMBH binaries in shaping the central structure of galaxies
depends not only on the efficiency of three-body interactions at hardening
the binary but also on how galaxy mergers affect the inner stellar density profiles,
i.e. on whether cores are preserved or steep cusps are regenerated during
major mergers. To bracket the uncertainties and explore  
different scenarios we run three different sets of Monte Carlo realizations.
In the first (`{\it cusp regeneration}') we assume, as in Paper I, that the 
stellar cusp $\propto 
r^{-2}$ is promptly regenerated after every major merger event, i.e., 
we replenish the mass displaced by the binary and reset $r_i=0$ after 
every major merger. In the second (`{\it core preservation}') the effect
of the hierarchy of SMBH binary interactions is instead cumulative, 
i.e., $r_i$ is allowed to grow continuously during the cosmic evolution 
of the host. This second model is supported by N-body simulations 
involving mergers of spherical galaxy models with different density 
profiles, and 
showing that the remnant profile is quite close to the profile of the 
progenitors -- in other words that the core appears to be preserved 
during such mergers (e.g. Fulton \& Barnes 2001).
The third set is similar to the second, except that stellar ejection is 
switched off at separation $a=0.1\,a_h$, as may be expected in the case the 
depletion of the loss cone were to `stall' the binary around that separation 
(e.g. Milosavljevic \& Merritt 2002). Thereafter the binary is assumed to
shrink rapidly due to (say) gas processes and coalesce. As shown below, this 
case yields results that are intermediate between the first two.   

Let us focus on the first two sets of realizations.
Figure \ref{fig1} shows how the mean core radius grows as a function of 
redshift in the simulated merger history of a $M_0=2\times 10^{12}\,\msun$ 
and a $M_0=10^{13}\,\msun$ 
halo. The cores created by shrinking SMBH binaries tend to remain small
($r_c<1\,$pc) until relatively recent times: on average, core radii approximately
double between $z=1$ and the present epoch. Core radii are also twice as large 
in the `core preservation' case than in the 
`cusp regeneration' one. (Note that in the latter averaging over all 
realizations smoothes out the $r_i=0$ resetting after every major 
merger event.) 

\begin{figurehere}
\vspace{0.4cm}
\centerline{
\psfig{file=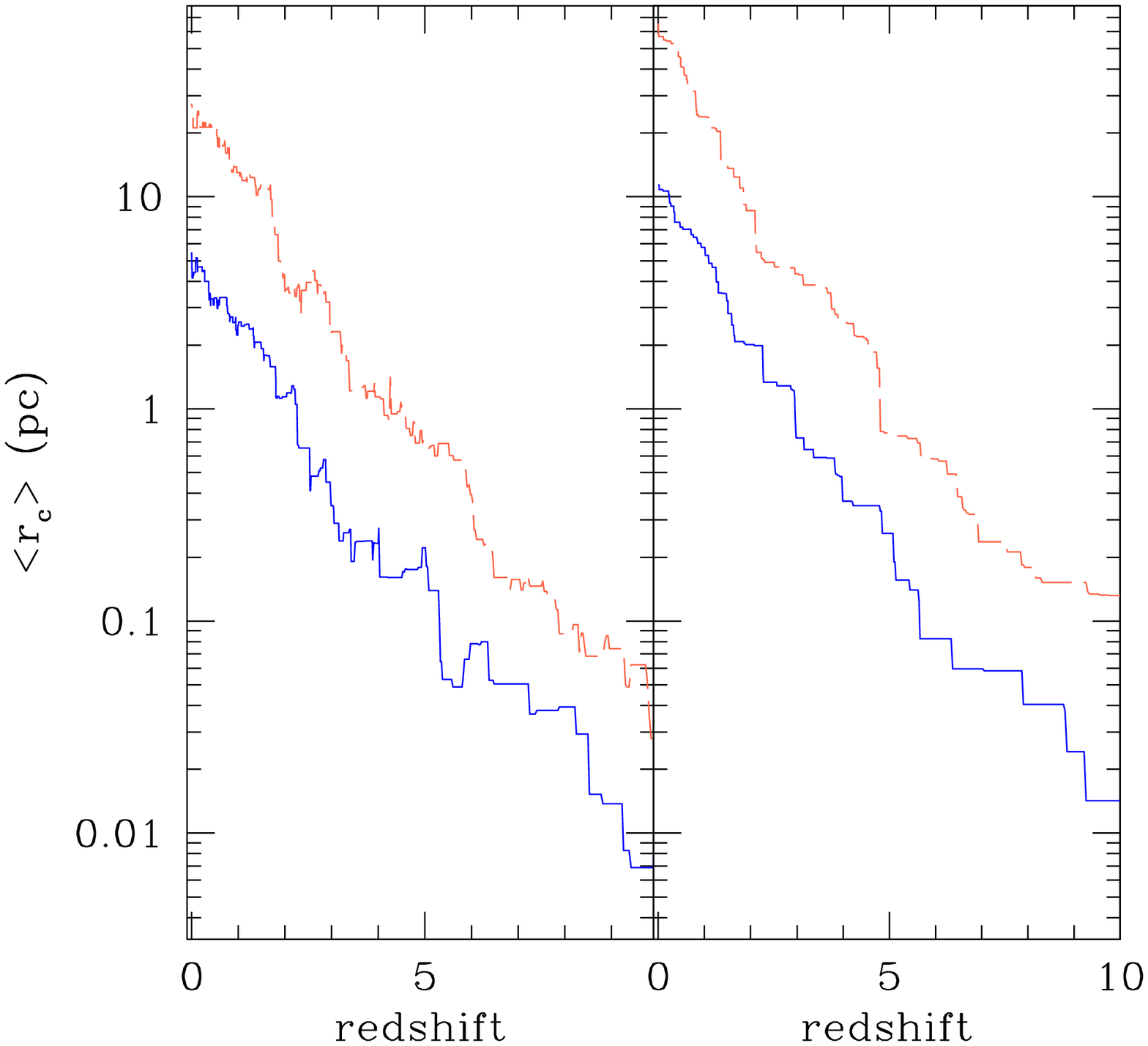,width=3.in}}
\caption{\footnotesize Mean (averaged over 20 realizations) core radius
of the main halo of the merger tree as a function of redshift. The expansion
of the core is supported by the binding energy liberated by shrinking 
SMBH binaries.  
{\it Solid line:} $M_0=2\times 10^{12}\,\msun$ halo. {\it Dashed line:} 
$M_0=10^{13}\,\msun$ halo. {\it Left panel:} cusp regeneration case. 
{\it Right panel:} core preservation case.}
\label{fig1}
\end{figurehere}
\vspace{+0.4cm}

Because of the lower central densities, binaries that form in
the mergers of massive galaxies decay much less rapidly in the model that
preserves cores than in the one where cusps are regenerated.
Figure \ref{fig2} shows the expected mean core radius at $z=0$, as a 
function of halo mass. Small galaxies have tiny cores or no core at all, 
while massive galaxies have core radii that can exceed hundreds of parsecs.

This is due
to two effects: (1) our accretion recipe is fixed in order to reproduce
the observed local $m_{\rm BH}-\sigma_*$ relation. More massive halos 
have more massive central SMBHs and hence larger cores; and (2) the assumed 
`bias' in the frequency of primordial seed BHs. Low mass galaxies today are 
`antibiased' as they are assigned a smaller abundance of BH seeds with respect
to the mean at early epochs, and some of them never host binaries. By contrast,
more massive galaxies form through the merging of an above-average 
number of primordial high-$\sigma$ minihalos, and so the damage done by
shrinking BH binaries is enhanced.
Note that, in the case of core 
preservation, the core radius is typically $\sim 3$ times larger than the 
radius of the ``sphere of influence'' of the final SMBH, 
$r_{\rm BH}=Gm_{\rm BH}/\sigma_*^2$. The predicted $r_c$ of the 
most massive 
halos in the core preservation case is larger than the core one would 
simply compute from equation (\ref{rc}) assuming $r_i=0$, an equal mass BH
merger ($m_1=m_2$), and a final merged BH mass that satisfies the 
$m_{\rm BH}-\sigma_*$ relation of Ferrarese (2002), $r_c \propto m_{\rm BH}/
\sigma_*^2\propto m^{0.57}_{\rm BH}$.

How do our results compare with observations? The simple models for core 
creation
described above yields core radii that scale almost linearly with galaxy mass, 
$r_c\propto M_0^{0.8\div0.9}$ in the range $10^{12}<M_0<4\times10^{13}\,\msun$.
A similarly scaling relation was observed by Faber \etal (1997). Our core sizes, 
however, are sensitive to the adopted density profile. We have assumed here that the 
continuous eroding 
action of shrinking binaries generates a flat density core. In the simulations of
Milosavljevic \& Merritt (2001), mergers of equal-mass stellar systems containing SMBHs
and steep central density cusps produce nuclei with shallower cusps
$\propto r^{-1}$ rather than flat cores. 
\begin{figurehere}
\vspace{0.4cm}
\centerline{
\psfig{file=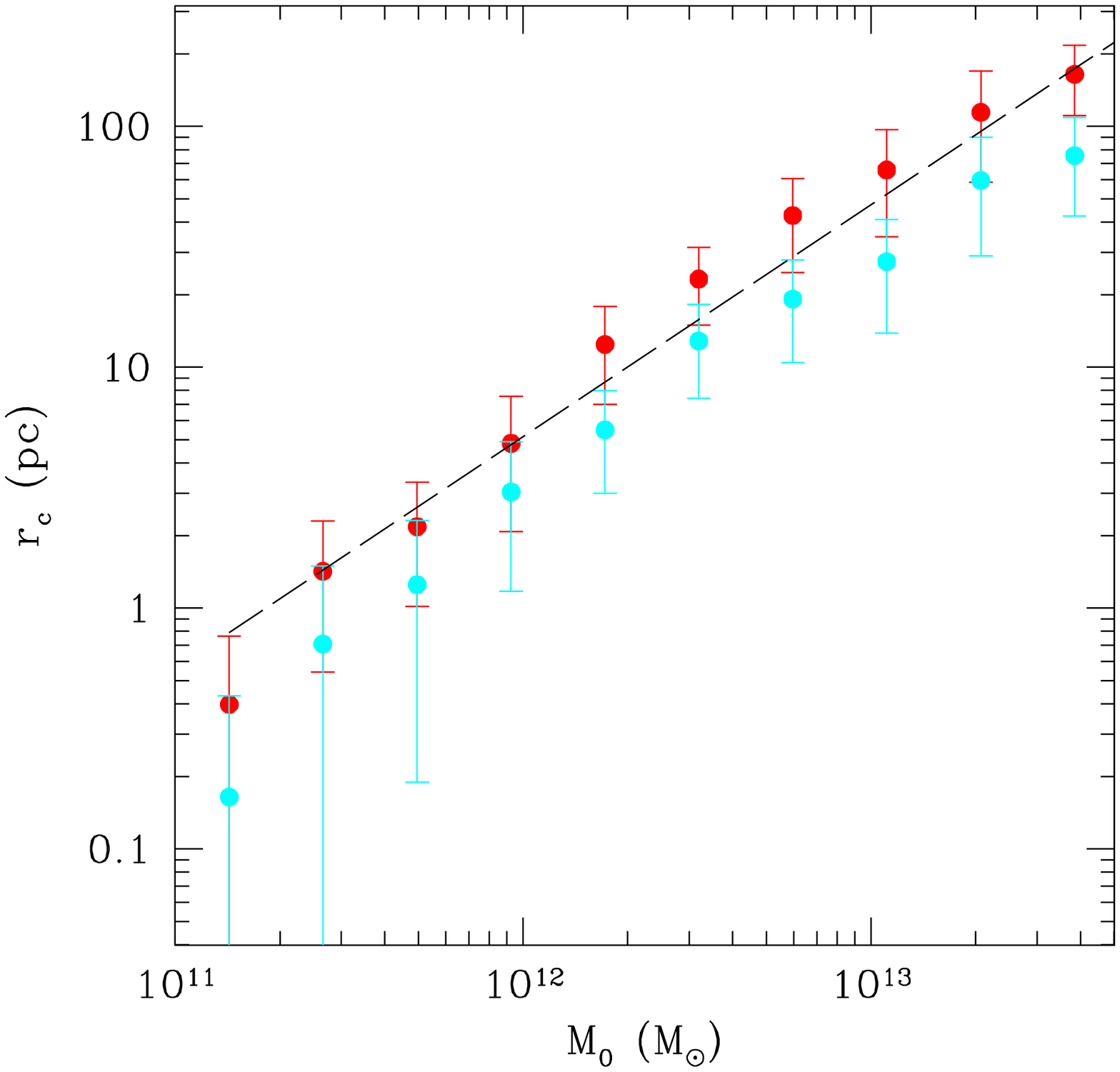,width=3.in}}
\caption{\footnotesize Mean core radius as a function of halo mass.
Errorbars are 1-$\sigma$ rms. {\it Lower dots:} cusp regeneration case. 
{\it Upper dots:} core preservation case. 
{\it Dashed line:} 
core radius computed from eq. (\ref{rc}) assuming $r_i=0$, an equal mass BH
binary ($m_1=m_2$), and a final merged BH mass that satisfies the 
$m_{\rm BH}-\sigma_*$ relation. Note that this prescription overpredicts
$r_c$ at low masses: this is due to small SMBHs deviating in our model
from the extrapolated $m_{\rm BH}-\sigma_*$ relation.} 
\label{fig2}
\end{figurehere}
\vspace{+0.4cm}
Observationally, even `core' galaxies exhibit very shallow power-law 
density profiles. A better test of our model predictions against galaxy data is provided
by the `mass deficit', i.e. the mass in stars that must be added to the observed cores 
to produce a stellar $r^{-2}$ cusp.\footnote{More precisely, the mass deficit 
is defined as the difference in integrated mass between the deprojected density 
profile and a $r^{-2}$ cusp extrapolated inward from the break radius.}
Milosavljevic \etal (2002) (see also Ravindranath, Ho, \& Filippenko 2002) 
have recently shown that the mass deficit 
inferred in a sample of early-type `core' galaxies 
correlates well with the mass of their nuclear black holes, consistent with the
prediction of coalescing SMBH binary models. In our model this quantity, 
\beq
M_{\rm def}=4\sigma_*^2r_c/(3G),
\eeq
is proportional to the total stellar mass ejected by shrinking binaries over cosmic history, and
is directly related to the hardening efficiency.

Figure \ref{fig3} compares the mass deficit inferred from the data with 
the same quantity found at $z=0$ in our merger tree,
\beq
M_{\rm def}/\msun=(8.2\pm3.8)\,(m_{\rm BH}/\msun)^{0.98\pm0.03}
\eeq
(core preservation) and
\beq
M_{\rm def}/\msun=(11.5\pm4.2)\,(m_{\rm BH}/\msun)^{0.93\pm0.02}
\eeq
(cusp regeneration), where the fit was performed in the same $m_{\rm BH}$ 
range of the observations. The cusp regeneration case predicts 
$M_{\rm def}\simeq 2.7 
m_{\rm BH}$ for $m_{\rm BH}=10^9\,\msun$, and clearly underestimates the mass 
deficit observed in massive `core' galaxies, $\langle M_{\rm def}\rangle
\simeq 10\,m_{\rm BH}$. We find that typically about 
half of $M_{\rm def}$ is generated in the last binary merger. 
Note that the mass deficit defined in this way is about 25\% larger than 
the stellar mass physically ejected in our simulations, since in our 
scheme $\sigma_*$ (hence the mass deficit) keeps increasing after binary
coalescence as a consequence of the accretion of small dark matter satellites
along the merger tree. 

\begin{figurehere}
\vspace{0.4cm}
\centerline{
\psfig{file=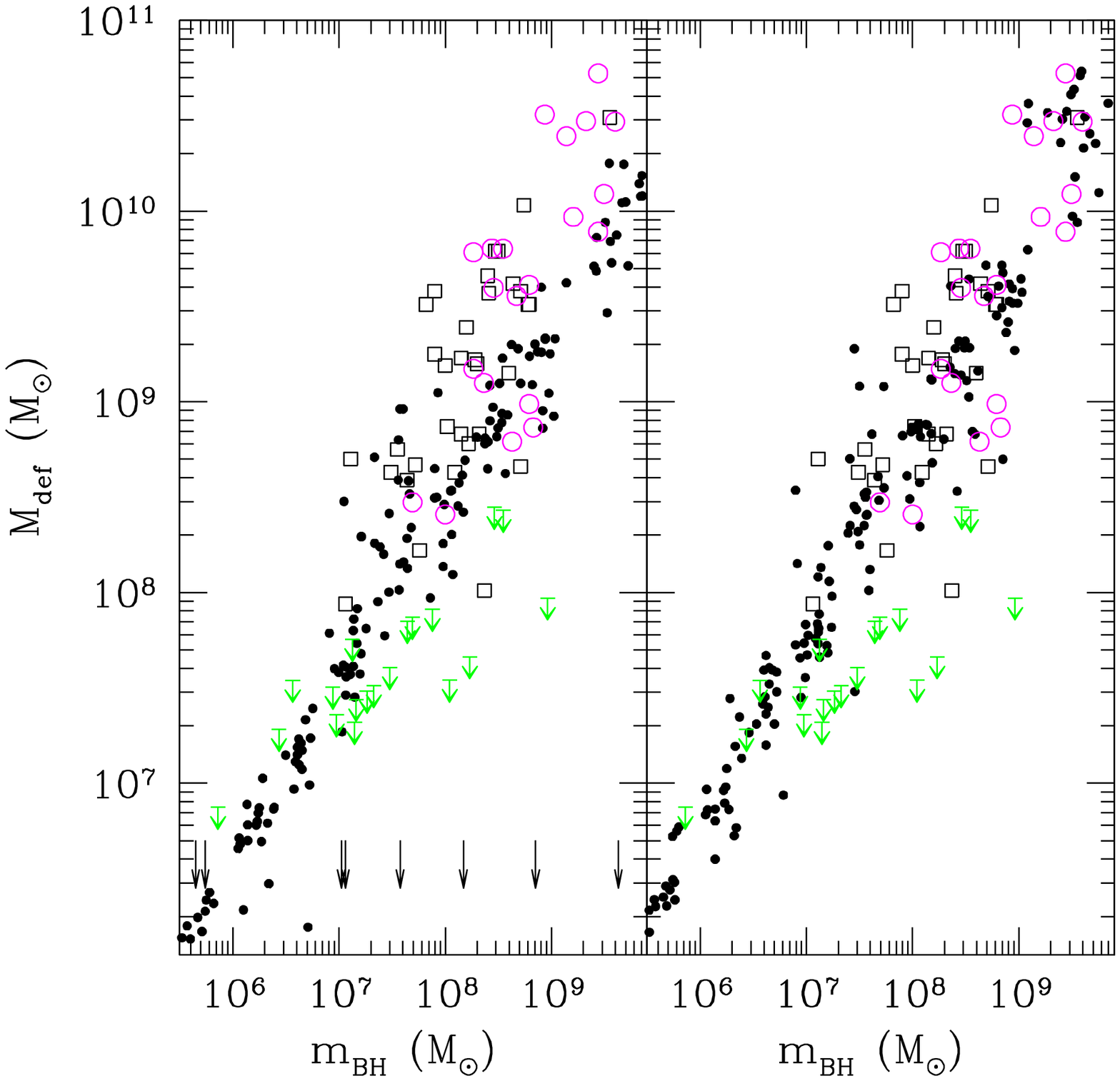,width=3.in}}
\caption{\footnotesize 
Mass deficit produced at $z=0$ by shrinking SMBHs 
in our merger tree as a function of nuclear SMBH mass ({\it filled dots}). 
{\it Left 
panel:} cusp regeneration case. {\it Right panel:} core preservation case. 
Galaxies without a core (i.e. those that have never
experienced a BH-BH merger or with their cusp recently regenerated) are
shown as vertical arrows at an arbitrary mass deficit of $10^6\msun$. 
{\it Empty squares}: mass deficit inferred in a sample of galaxies by 
Milosavljevic \etal (2002). {\it Empty circles}: same for the `core' galaxies 
of Faber \etal (1997) sample. {\it Upper limits} are shown for the 
`cuspy' galaxies of Faber \etal (1997) sample, assuming a flat core within
the upper limit on the core size.
} \label{fig3}
\end{figurehere}
\vspace{0.4cm}

The predicted $M_{\rm def}-m_{\rm BH}$ correlation 
is too tight compared to the observations: a distribution of inner density 
profiles before and after galaxies merge and observational errors may
both contribute to the large scatter in the data points. We also draw 
attention to the fact that the mass deficits estimated for the cuspy 
galaxies of Faber 
\etal (1997) should be strictly considered as upper limits, and the  
observed $M_{\rm def}-m_{\rm BH}$ relation may then be steeper than predicted 
(see also Ravindranath \etal 2002). A few galaxies in the sample of Faber \etal 
have cores significantly smaller 
than expected from their central BH mass (e.g. NGC3115, with a  break 
radius $\sim 2$ pc and $m_{\rm BH}=10^{9}\,\msun$). 
Similarly the Milky Way galaxy has a BH with $m_{\rm BH}=3\times 
10^{6}\,\msun$ and a stellar core radius $r_c=0.38\,$pc  (Genzel \etal
2000). Within our simple scheme,
objects with these properties are very rare in the `core preservation' 
case, and may require some form of cusp regeneration mechanism.  

The mass deficit measured in our third set of realizations is depicted in Figure
\ref{fig4}. This assumes core preservation and switches off stellar ejection 
when $a=0.1\,a_h$, i.e. stellar dynamical processes become inefficient at 
small binary separations. Even in this scenario stellar cusps may 
be efficiently destroyed over cosmic time by decaying SMBH binaries:
the expected mass deficit is less than a factor of 2 smaller than in 
the standard core preservation case. 
\begin{figurehere}
\vspace{0.4cm}
\centerline{
\psfig{file=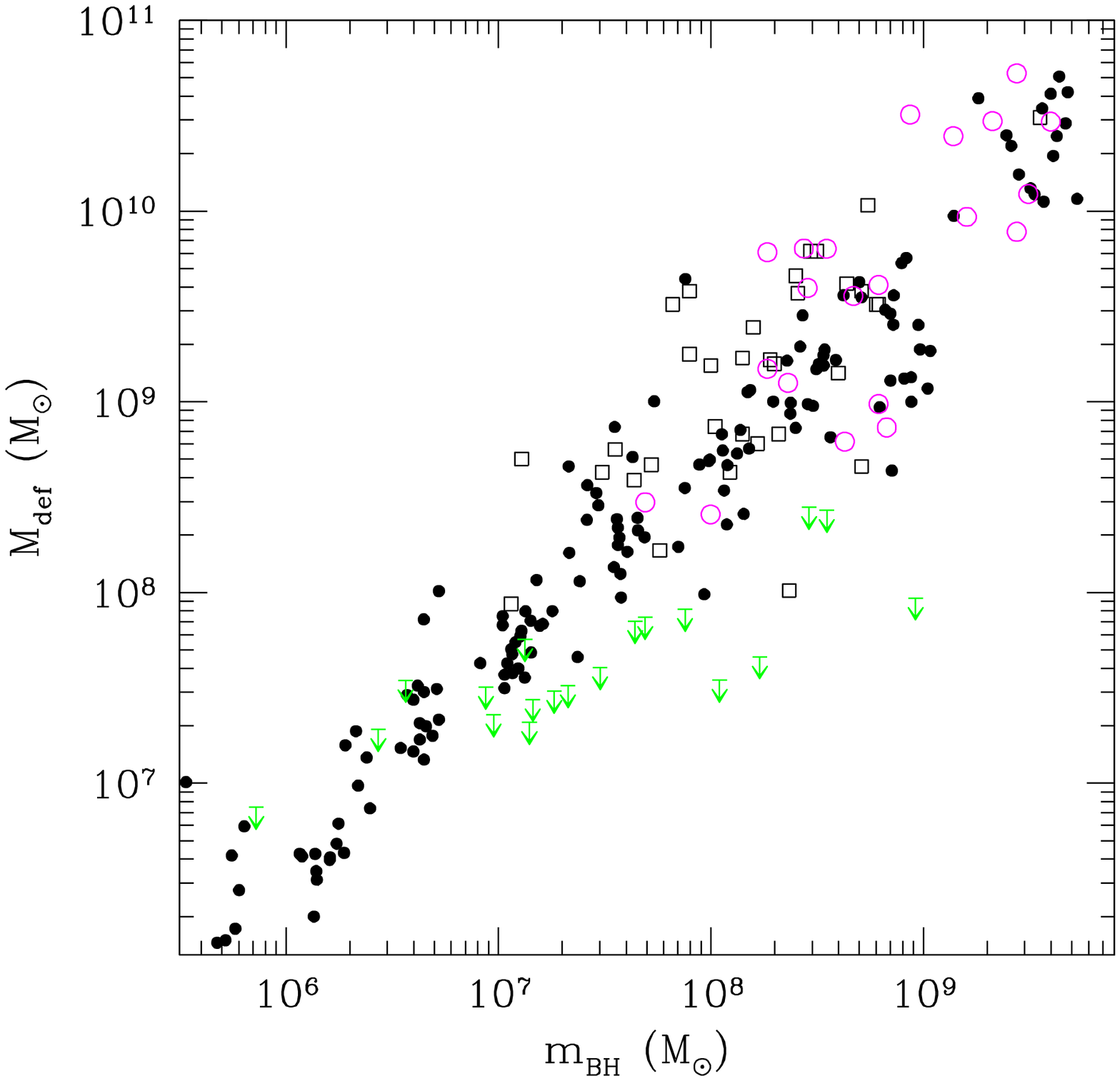,width=3in}}
\caption{\footnotesize Same as Fig. \ref{fig3}, except that stellar ejection is 
switched off at separation $a=0.1\,a_h$; thereafter the binary is assumed to
shrink rapidly due to gas processes and coalesce. The realizations 
assume cores are preserved during galay mergers.} 
\label{fig4}
\end{figurehere}
\vspace{+0.4cm}

\section{Discussion}

We have assessed a hierarchical structure formation scenario in which 
galaxy inner cores are created from the binding energy liberated by 
shrinking SMBH binaries. The binary orbital decay heats the surrounding 
stars, eroding a preexisting stellar cusp $\propto r^{-2}$. 
A simple model in which the effect of the hierarchy
of SMBH interactions is cumulative produces a correlation between `mass 
deficit' and the mass of nuclear SMBHs with a normalization and slope that are
comparable to the observed relation. This model is also able to reproduce
the observed scaling relation between galaxy luminosity and core size. 
Relating the dark halo mass to galaxy luminosity using the mass-to-light 
ratios of van den Bosch, Mo, \& Yang (2002), we find $r_c\propto L_V^{1.2}$ in the mass
range $10^{12}<M_0<4\times10^{13}\,\msun$, in agreement with  
the scaling observed by Faber \etal (1997). Despite these encouraging 
results, models of core formation by binary SMBHs remain uncertain. 
The ability of SMBH binaries in modifying the inner density profiles of 
galaxies depends on whether cores are preserved or steep cusps are 
regenerated during galaxy mergers, and on the poorly known efficiency of 
three-body interactions at hardening the binary. 

There are of course further complications: (1) If the age of the stellar 
core$+$BH system is much larger than the local relaxation time, 
then an equilibrium distribution of bound stars will be set up 
within the BH sphere of influence. We find that such a steady-state density cusp, 
$\rho_*\propto r^{-7/4}$ (Peebles 1972; Bahcall \& Wolf 1976; Frank \& 
Rees 1976; Lightman \& Shapiro 1977) will have no
time to develop in our hierarchical scheme for the assembly of SMBHs, as
the relaxation timescales are typically rather long.
If the age of the system is instead much less than the relaxation time,
and the hole grows by accreting gas on a time scale long compared with the
orbital period of the surrounding stars, then again a power-law density 
cusp will be created, this time with slope $r^{-3/2}$ extending 
to $r_{\rm BH}$ 
(Young 1980). As the mass of the hole increases, the core will start to
loose its identity until, when $r_{\rm BH}\gta r_c$, the cusp will join 
smoothly onto the $r^{-2}$ isothermal profile. Adiabatic growth of stellar 
cusps is typically not important within our scheme, since we find 
that the core mass $M_c$ exceeds $m_{\rm BH}$, i.e. that $r_{\rm BH}<r_c$ 
in most systems; (2) Dark matter particles will be ejected by shrinking  
SMBH binaries in the same way as the stars, i.e. through the gravitational
slingshot. Their contribution to binary hardening will be weighted by 
the fraction $f_{\rm DM}=\rho_{\rm DM}/(\rho_{\rm DM}+\rho_*)$, typically 
$\ll 1$ in galaxy 
cores for our assumed density distributions. The NFW profile is shallower than 
the SIS profile in the inner regions, so ejection of a given mass will 
create a larger core in an NFW profile. The mass ejected in dark matter is, 
however, significantly less than the mass ejected in stars. Eventually, 
the destruction of dark matter cusps by binary SMBHs may lead to cores of 
the same extent as the stellar ones. Furthermore, the erosion of the 
stellar cusp may lower the central dark matter density through a 
process (`adiabatic expansion') that is the opposite of the adiabatic 
compression of dark halos in response to baryonic cooling and infall 
(Blumenthal \etal 1986). While the observed rotation curves of
dwarf and lower surface brightness galaxies suggest that the inner
regions have constant density cores rather than the density cusps 
predicted by CDM (McGaugh \& De Blok 1998; Swaters \etal 2003), 
shrinking BH binaries will not alleviate significantly this 
`CDM cusp problem', since the predicted mass deficit is much lower than
that required to fit the rotation curves. 

\acknowledgements
We have benefitted from discussions with G. Blumenthal, S. Faber, M. 
Haehnelt, M. Milosavljevic, and M. Rees. Support for this work was provided 
by NASA through grant NAG5-11513 and by NSF grant AST-0205738 (P.M.).

{}

\end{document}